\documentclass[12pt]{article}
\usepackage{epsfig, amssymb}
\usepackage{graphicx,epsfig}
\begin{document}
\begin{titlepage}
\thispagestyle{empty}
\begin{flushright}
\end{flushright}

\bigskip

\begin{center}
\noindent{\Large \textbf
{Stochastic quantization and holographic Wilsonian renormalization group of massless fermions in AdS}}\\
\vspace{2cm} \noindent{
Jae-Hyuk Oh\footnote{e-mail:jack.jaehyuk.oh@gmail.com}}

\vspace{1cm}
  {\it
Department of Physics, Hanyang University, Seoul 133-791, Korea\\
 }
\end{center}

\vspace{0.3cm}
\begin{abstract}
We have studied holographic Wilsonian renormalization group
(HWRG) of free massless fermionic fields in AdS space and its stochastic quantization(SQ) 
by identifying the Euclidean action with its boundary on-shell action.
The natural extension of the relation between stochastic 2-point correlation function and double trace coupling obtained from HWRG computation conjectured in arXiv:1209.2242 to fermionic fields is established. We have confirmed that the stochastic 2-point function precisely captures the radial flow of the double trace coupling via this relation.
\end{abstract}
\end{titlepage}

\newpage

\tableofcontents
\section{Introduction}

A year ago, one of the authors in this paper and Dileep P. Jatkar published several
papers\cite{Oh:2012bx,Jatkar:2013uga}, in which they discussed
concrete mathematical relationships between stochastic quantization (SQ)\cite{Wu1,Huffel1,Dijkgraaf1} 
and holographic Wilsonian renormalization group(HWRG)\cite{Polchinski1,Hong1}
\footnote{There are some of studies to relate SQ and AdS/CFT\cite{Petkou1,Minic:2010pw} in different directions from ours.}. 
They have pointed out that the double trace deformations in the radial evolution of the boundary effective action, a solution of Hamilton-Jacobi type equations derived from certain free theories defined in AdS space can be completely captured by stochastic 2-point correlations.  

HWRG is a process to compute radial evolutions of boundary effective theories, whose dual is certain 
bulk gravity theories in AdS space. The boundary theory is defined on $r=\epsilon$ hyper surface,
where $r$ is radial coordinate of AdS space and $\epsilon$ is its arbitrary cut-off. 
By requesting that total theory which contains both the bulk gravity action and the boundary effective 
action does not depend on the cut-off $\epsilon$, a Schroedinger type equation is obtained which describes the radial evolution of the boundary effective action, and especially in classical limit, it is reduced to Hamilton-Jacobi equation.

On the other hand, SQ describes stochastic time evolution of a non-equilibrium thermodynamic system 
contacting with its large thermal reservoir(with definite temperature). We start with a certain theory $S_c$ 
defined in Euclidean spacetime at $t=t_0$, where $S_c$ is called Euclidean action, $t$ is stochastic time and $t_0$ is the initial value of it. Then, the system is evolving into the thermal 
equilibrium with its surrounding as the stochastic time goes by. One can recognize that the thermodynamic partition function in this 
equilibrium state is (at least mathematically) the same with Euclidean partition function with $S_c$. 
This relaxation process is described by a diffusion equation called Langevin equation as well as it is 
done by a Schroedinger type equation called Fokker-Planck equation.

In \cite{Oh:2012bx,Jatkar:2013uga}, the authors argued that the two different Schroedinger 
type equations appearing in SQ and HWRG can be identified in the case of certain bulk theories 
once the stochastic time $t$ is identified to the radial coordinate of AdS space $r$ and the Euclidean 
action $S_c$ is given by $S_c=-2I_{os}$, where $I_{os}$ is called on-shell action, which is obtained from 
the bulk action by substituting regular solutions of bulk equations of motion into it. 

Such tries were quite successful in that they have provided several explicit examples such as massless scalar field theory in AdS$_2$, U(1) gauge 
theory in AdS$_4$ and conformally coupled scalar field theory in AdS$_4$ and all the radial 
evolutions of their double trace deformations in the boundary effective actions are completely 
captured by stochastic two point correlation functions via the following relation:
\begin{equation}
\label{core-boson}
\langle \phi_i(k,t) \phi_j(-k,t) \rangle_H^{-1} = \langle \phi_i(k,t) \phi_j(-k,t) \rangle_S^{-1}-\frac{1}{2} \frac{\delta^2 S_c}{\delta \phi_i(k,t)\delta \phi_j(-k,t)},
\end{equation} 
where $\phi_i$ represent any stochastic(bosonic) fields and the indices, $i$ and $j$ 
denote collections of all the indices that the fields $\phi$ carry. 

Fokker-Planck approach also yields the same result with what Langevin equation gives. 
The authors in \cite{Oh:2012bx,Jatkar:2013uga} have found that the double trace deformation parts 
of the boundary effective actions for the examples that they provided are precisely obtained from Fokker-Planck Lagrangian density through the relation as
\begin{equation}
\label{the-relation=}
S_B=\int^t_{t_0} dt^\prime \int d^d k \mathcal L_{FP}(\phi(k,t^\prime)_i),
\end{equation}
where $\mathcal L_{FP}$ is Fokker-Planck Lagrangian density\footnote{The Fokker-Planck Lagrangian density is derived from stochastic partition function, which will be discussed in section \ref{Stochastic quantization of massless fermions in AdS} in detail.}.

In this paper, we extend these studies to massless bulk fermions defined in AdS${_{d+1}}$. 
Massless fermions in $d+1$-dimensional AdS spacetime have two different merits as follows. 
Firstly, it allows both standard and alternative quantizations
\cite{Daniel1,Witten11,Klebanov:1999tb,Witten:2001ua,Jatkar:2012mm,Ioannis1,Sebastian1,
Sebastian2,Sebastian3}. In the mass range of bulk fermions as $0\leq m <\frac{1}{2}$ 
in $d+1$ dimensional AdS space, it allows alternative quantization.
Therefore, it shows two different fixed points which correspond to standard and alternative quantization 
schemes in the boundary CFT's and present diverse renormalization group flows 
from one fixed point to another and many other curves which connect various points which 
are not fixed points.
Secondly, by using an appropriate field redefinition, it maps to fermionic action effectively 
defined in $\mathbb R^{d+1}_+$, a half of flat, $d+1$-dimensional spacetime, (since $r$ is semi-infinite) with zero mass. This makes HWRG and SQ computations much simpler.

In section \ref{section2}, we review HWRG of massless bulk fermion in AdS${_{d+1}}$. 
This computation is nicely performed in \cite{Elander:2011vh,Laia:2011wf} already and 
various renormalization group curves are studied therein. Especially, the authors in \cite{Elander:2011vh} performed the appropriate redefinition of bulk fermionic fields as 
$\psi=(\sqrt{g}g^{rr})^{1/4}\Phi$, where $g$ is bulk metic determinant, 
$g^{rr}$ is inverse of $g_{rr}$, 
$rr$-component of the bulk metric, $\psi$ is fermionic field defined in $AdS$ space and $\Phi$ is the fermionic in $\mathbb R^{d+1}_+$ with $r$ dependent mass term. 
If one requests $m=0$, then the theory becomes massless fermion in $\mathbb R^{d+1}_+$. Thanks to this fact, one can easily compute the radial evolution of its boundary effective action, especially its double trace deformation term, which is given by
\begin{equation}
\label{double-tr}
D_{\alpha\beta}(\epsilon,k)=-\frac{E_\alpha(k)}{|k|}\delta_{\alpha\beta}\frac{\sinh(|k|\epsilon)+\Delta_\alpha\cosh(|k|\epsilon)}{\cosh(|k|\epsilon)+\Delta_\alpha\sinh(|k|\epsilon)}
\end{equation}
where $\alpha$, $\beta$ are spinor indices of fermionic field $\chi_+$($\chi_-$) and these indices are
not summed in the above expression.
$\chi_+$($\chi_-$) is upper(lower) half components of the fermionic field(spinor) $\Phi$ when $d$ is odd, i.e.
\begin{equation}
\Phi= \left(\begin{array}{cc} \chi_+ \\ \chi_-\end{array}\right).
\end{equation}
When $d$ is even, 
\begin{equation}
 \chi_\pm=\frac{1\pm\Gamma^{\hat r}}{2}\Phi,
\end{equation}
where $\Gamma^{\hat r}\equiv\gamma^{\hat 0}\gamma^{\hat 1}...\gamma^{\hat {d-1}}=\frac{1}{d!}\varepsilon_{\mu\nu\rho...}\gamma^{\hat\mu}\gamma^{\hat\nu}\gamma^{\hat\rho}...$, $\varepsilon_{\mu\nu\rho...}$ is fully antisymmetric tensor
and $\gamma^{\hat\mu}$ 
is $\gamma$-matrices on the boundary of (Euclidean)AdS$_{d+1}$, which are labeled by the spinor indices $\alpha$ and $\beta$.
$E_\alpha(k)$ are Eigen values\footnote{For example, the Eigen values of hermitian matrix $k_\mu\gamma^\mu$ in 4-dimensional Euclidean spacetime are given in Eq(\ref{eigen-values-of-k-gamma}).}
of hermitian matrix $k_\mu\gamma^{\hat\mu}$ and $\Delta_\alpha$ is a constant. 
$k_\mu$ is $d$-dimensional momenta along AdS boundary directions.

It turns out that this result is precisely captured by 
stochastic quantization of Euclidean action $S_c$, 
which is given by the boundary on-shell action $I_{os}$ of the bulk fermions $\chi_+$ and $\bar\chi_+$, evaluated  at 
$r=\epsilon$ hyper surface via the suggested relation $S_c=-2I_{os}$ in \cite{Oh:2012bx,Jatkar:2013uga}.
We have discussed the exact process of this in section 
\ref{Stochastic quantization of massless fermions in AdS} and 
have found that the double trace coupling(\ref{double-tr}) is obtained from the fermionic extension of the 
relation(\ref{core-boson}), which is given by
\begin{equation}
\langle \chi_\alpha(k,t) \bar \chi_\beta(k,t) \rangle_H^{-1} = \langle \chi_\alpha(k,t) \bar \chi_\beta(k,t) \rangle_S^{-1}-\frac{1}{2} \frac{\overrightarrow\delta}{\delta \bar \chi_\alpha(-k,t)} 
S_c \frac{\overleftarrow\delta}{\delta \chi_\beta(-k,t)},
\end{equation}
where $\langle \chi_\alpha(k,t) \bar \chi_\beta(k,t) \rangle_H$ 
is double trace coupling and $\langle \chi_\alpha(k,t) \bar \chi_\beta(k,t) \rangle_S$ 
is stochastic 2-point correlator of the fermion $\chi_+$ and $\bar\chi_+$. 
The arrows on the functional differentiations indicate the directions that the variations act on.

We also evaluate Fokker-Planck action in classical limit and get a 
boundary effective action via the relation(\ref{the-relation=}). 
It turns out that the double trace deformation part of the boundary effective action is precisely 
recovered via the relation.

\section{Holographic Wilsonian renormalization group of massless fermions in AdS$_{d+1}$}
\label{section2}
In this section, we will review holographic Wilsonian renormalization group for massless fermions defined in AdS$_{d+1}$
\footnote{Detailed discussion about boundary effective action for fermions appear in \cite{Elander:2011vh,Laia:2011wf,Iqbal:2009fd}.}.
\subsection{Holographic set up of the bulk fermions}
We start with the action 
\begin{equation}
\label{fermionic-action}
S_{bulk}=-i\int drd^dx \sqrt{g}\left( \bar \psi \Gamma^{M}\nabla_M \psi - m \bar \psi \psi\right),
\end{equation}
where 
\begin{equation}
\nabla_N= \partial_N + \frac{1}{4}\omega_N^{\hat A\hat B} \Gamma_{\hat A \hat B}.
\end{equation}
$M$ and $N$ are spacetime indices whereas $\hat A$ and $\hat B$ are tangent space indices all of which run from 1 to $d+1$ and the spacetime metric represents Euclidean $d+1$ dimensional 
AdS space, given by
\begin{equation}
\label{background-metric}
ds^2=\frac{1}{r^2}\left( dr^2 + \sum_{\mu,\nu=1}^d\delta_{\mu\nu}dx^\mu dx^\nu \right),
\end{equation}
where $\mu$ and $\nu$ are boundary spacetime indices, which run from 1 to $d$. $m$ is mass of the bulk fermions and we set it to be zero in the following discussion. 
$\omega_N^{\hat A\hat B}=e^{\hat A}_M\nabla_N e^{\hat B\ M}$ is connection 1-form and $e^{\hat A}_M$ is vielbein.

In our further discussion, we restrict ourselves in massless fermionic fields in the bulk.
Massless bulk fermions in AdS space contain some of merits. Firstly, if their masses are in the range of $0\leq m < \frac{1}{2}$, then all the classical solutions of the bulk fermions
become normalizable and both standard and alternative 
quantizations in dual CFT are allowed. Since these two different quantization schemes are possible, 
we have two different $UV$ and $IR$ fixed points in dual CFT defined on the conformal boundary and various 
RG flows connecting those $UV$ and $IR$ fixed points exist. Another interesting observation on these fermions\cite{Elander:2011vh} is that if one defines a new fermionic field $\Phi$ and $\bar \Phi$ such as
$\Phi=(gg^{rr})^{1/4} \psi$ and $\bar\Phi=(gg^{rr})^{1/4} \bar\psi$, then the primitive action(\ref{fermionic-action}) is able to be given by
\begin{equation}
\label{Phi-fermionic-action}
S_{bulk}=-i\int dr d^d x \left( \bar\Phi \Gamma^{\hat r} \partial_r \Phi + \bar\Phi \Gamma^{\hat \mu} \partial_\mu \Phi  - \sqrt{g^{rr}}m\bar\Phi\Phi\right),
\end{equation}
where $g^{rr}=r^2$, the inverse of $rr$-component metric and $g=det(g_{MN})$.

If dimensionality of the boundary spacetime is odd, the gamma-matrices $\Gamma^{M}$ are given by
\begin{eqnarray}
 \Gamma^{\hat r}=\left(\begin{array}{cc} \mathbb I & 0 \\ 0 & -\mathbb I\end{array}\right) {,\ \ }\Gamma^{\hat\mu}=\left(\begin{array}{cc} 0 & \gamma^{\hat \mu} \\ \gamma^{\hat \mu}& 0\end{array}\right) 
{,\rm \ and\ } \Phi=\left(\begin{array}{c} \chi_+ \\ \chi_-\end{array}\right),
\end{eqnarray}
where $\gamma^{\hat \mu}$ are the gamma-matrices in $d$-dimensional boundary spacetime and $\mathbb I$ 
is the identity operator being the same size with the boundary 
gamma-matrices, $\gamma^{\hat \mu}$.
$\chi_+$ and $\chi_-$ are Dirac spinors in the sense of the boundary but they are two Weyl fermions in the bulk.
$\bar\Phi$ is related to $\Phi$ by imposing hermicity of the bulk action, then it is given by
\begin{equation}
\label{psi-decomposition}
\bar\Phi=\Phi^{\dagger}\Gamma^{\hat 0}=(\bar\chi_-{\ } \bar\chi_+),
\end{equation}
where $\bar\chi_\pm=\chi^\dagger_\pm \gamma^{\hat0}$.

When the boundary spacetime has even dimensionality, $\Gamma^{\hat r}$ is $\gamma^{\hat{d+1}}\equiv\gamma^{\hat 0}\gamma^{\hat 1}...\gamma^{\hat {d-1}}$ and $\Gamma^{\hat\mu}=\gamma^{\hat\mu}$. 
In this case, the bulk fermion, $\Phi$ is Dirac fermion and since the boundary spacetime is even, 
it can be decomposed into two Weyl fermions
on the boundary as
\begin{equation}
\chi_\pm=\frac{1\pm\Gamma^{\hat r}}{2}\Phi {\rm \ \ and \ }\bar\chi_\pm=\bar\Phi\frac{1\mp\Gamma^{\hat r}}{2}.
\end{equation}
%
%
%
%
The action(\ref{Phi-fermionic-action}) can be written in terms of
$\chi_\pm$ as
\begin{eqnarray}
\label{action-with-chi}
S&=&\left.-i\int dr d^dk \right[-\bar \chi_{+}(k)\partial_r \chi_-(-k) + \bar\chi_-(k)\partial_r \chi_{+}(-k) + \bar\chi_+(k)\gamma^{\hat \mu}(-ik_{\mu})\chi_+(-k) \\ \nonumber
&+& \bar\chi_-(k)\gamma^{\hat \mu}(-ik_{\mu})\chi_-(-k)].
\end{eqnarray}

\paragraph{Equations of motion and their solutions}
Varying the action(\ref{action-with-chi}) provides the following set of equations:
\begin{eqnarray}
\label{eq-chi+}
0&=&\partial_r \chi_+(-k)+\gamma^{\hat \mu}(-ik_{\mu})\chi_-(-k), \\
\label{eq-chi-}
0&=&-\partial_r \chi_-(-k)+\gamma^{\hat \mu}(-ik_{\mu})\chi_+(-k), \\
\label{eq-chibar+}
0&=&-\partial_r \bar\chi_-(k)+\bar\chi_+(k)\gamma^{\hat \mu}(-ik_{\mu}), \\
\label{eq-chibar-}
0&=&\partial_r \bar\chi_+(k)+\bar\chi_-(k)\gamma^{\hat \mu}(-ik_{\mu}),
\end{eqnarray}
The solutions $\chi_\pm$ are obtained by combining (\ref{eq-chi+}) and (\ref{eq-chi-}). A little manipulation on (\ref{eq-chi+}) provides
\begin{equation}
\label{manipulation-of-eq-chi+}
\chi_-(-k)=\frac{-ik_{\nu}\gamma^{\hat \nu}}{k^2} \partial_r \chi_+(-k),
\end{equation}
where we have used $\gamma^{\hat\mu}\gamma^{\hat\nu}k_\mu k_\nu=k^2\mathbb I$.
We plug this into (\ref{eq-chi-}) and get an equation in terms of $\chi_+$ only as
\begin{equation}
\label{chi+-equation-only}
\partial^2_r\chi_{\alpha,+}(k)-k^{2}\chi_{\alpha,+}(k)=0,
\end{equation}
where the subscript $\alpha$ is spinor index.
The most general solution of (\ref{chi+-equation-only}) is given by
\begin{equation}
\label{sol-most-chi+}
\chi_{\alpha,+}(k,r)=\chi^{(0)}_{\alpha,+}(k)\cosh(|k|r)+\chi^{(1)}_{\alpha,+}(k)\sinh(|k|r),
\end{equation}
where $\chi^{(0)}_{\alpha,+}(k)$ and $\chi^{(1)}_{\alpha,+}(k)$ are arbitrary spinors depending on momentum $k_{\mu}$ only.

We note that the spinor index $\alpha$ runs from 1 to $2^{\frac{d}{2}}$ when $d$ is even whereas it runs from $1$ to $2^{\frac{d-1}{2}}$ when $d$ is odd.
$k_\mu \gamma^\mu$ is hermitian matrix since all the $\gamma$-matrices are defined in Euclidean spacetime, so it can be diagonalized. 
Then, the matrix $k_\mu \gamma^\mu$ can have a form of
\begin{equation}
[k_\mu\gamma^\mu]_{\alpha\beta}=E_{\alpha}(k)\delta_{\alpha\beta},
\end{equation}
when $k_\mu \gamma^\mu$ is diagonalized and $E_{\alpha}(k)$ are its Eigen values.
Once it is diagonalized, Eigen kets of this matrix form a complete orthogonal basis. 
Therefore, another way to represent the solution(\ref{sol-most-chi+}) is as
\begin{eqnarray}
\label{sol-most-chi+++}
\chi_{\alpha,+}(k,r) &=&N_{\alpha,+}
\left(\begin{array}{cccc}0  \\ 0... \\{\cosh(|k|r)+\Delta_{\alpha,+}(k)\sinh(|k|r) } \\... 0 \\ 0  \end{array}\right) \leftarrow {\rm\ \alpha th\ row}, \\ \nonumber
&\equiv& N_{\alpha,+} [\cosh(|k|r)+\Delta_{\alpha,+}(k)\sinh(|k|r) ]  |\alpha\rangle,
\end{eqnarray} 
which denotes that its $\alpha$th row is non zero only. $|\alpha\rangle$ is given by
\begin{equation}
|\alpha\rangle=
\left(\begin{array}{cccc}0  \\ 0... \\1\\... 0 \\ 0  \end{array}\right) \leftarrow {\rm\ \alpha th\ row}
\end{equation}
and this ket is Eigen-vector providing Eigen value $E_\alpha(k)$. $N_{\alpha,+}$ 
is a normalization factor and $\Delta_{\alpha,+}(k)$ is an arbitrary momentum dependent constant.
We will use this basis to construct the radial evolution of double trace coupling of bulk massless fermion.

The field $\chi_-$ satisfies the same Klein-Gordon type equation and has its solution 
with different coefficients as
\begin{equation}
\label{sol-most-chi----}
\chi_{\alpha,-}(k,r)=\chi^{(0)}_{\alpha,-}(k)\cosh(|k|r)+\chi^{(1)}_{\alpha,-}(k)\sinh(|k|r),
\end{equation}
or by using the above basis, it is given by
\begin{equation}
\label{sol-most-chi-}
\chi_{\alpha,-}(k,r) =N_{\alpha,-}({\cosh(|k|r)+\Delta_{\alpha,-}(k)\sinh(|k|r) })|\alpha\rangle
\end{equation}
where $\chi^{(0)}_{\alpha,-}$ and $\chi^{(1)}_{\alpha,-}$ are arbitrary boundary momentum dependent spinors.
$N_{\alpha,-}$ is a normalization factor and $\Delta_{\alpha,-}(k)$ is an arbitrary momentum dependent constant.
One can apply the same technique to solve (\ref{eq-chibar+}) and (\ref{eq-chibar-}). We just list their solutions as
\begin{eqnarray}
\label{eq--+++}
\bar\chi_{\alpha,+}(k,r)&=&\bar\chi^{(0)}_{\alpha,+}(k)\cosh(|k|r)+\bar\chi^{(1)}_{\alpha,+}(k)\sinh(|k|r) \\ 
\label{eq+++--}
{\rm and \ \ }
\bar\chi_{\alpha,-}(k,r)&=&\bar\chi^{(0)}_{\alpha,-}(k)\cosh(|k|r)+\bar\chi^{(1)}_{\alpha,-}(k)\sinh(|k|r),
\end{eqnarray}
where again the spinors $\bar\chi^{(0)}_+(k)$, $\bar\chi^{(1)}_+(k)$ and $\bar\chi^{(0)}_-(k)$, $\bar\chi^{(1)}_-(k)$ are related one another as
\begin{equation}
\bar\chi^{(0)}_+(k)=-\frac{ik_\nu }{|k|} \bar\chi^{(1)}_-(k)\gamma^{\hat\nu}
{\rm\ \ and \ \ } \bar\chi^{(1)}_+(k)=-\frac{ik_\nu }{|k|}\bar\chi^{(0)}_-(k) \gamma^{\hat\nu},
\end{equation}
where the spinor indices are suppressed.

\subsection{Radial evolution of double trace operator for the bulk massless fermions}
\paragraph{Adding boundary terms for well posed variational problem\cite{Iqbal:2009fd}}
The bulk solutions, $\chi_+$, $\chi_-$ and $\bar \chi_+$, $\bar\chi_-$
are not completely independent but they are related respectively each other via their equations of motion (\ref{eq-chi+})-(\ref{eq-chibar-}). 
It turns out that such constraints raise sickness in the bulk action(\ref{fermionic-action}). If the bulk action, $S_{bulk}$ is varied with respect to the bulk fields, then it becomes
\begin{equation}
\delta S_{bulk}=i\left.\left.\int_{r=\epsilon} d^dk \right[ \bar \chi_+(k)\delta \chi_-(-k) + \bar \chi_-(k)\delta \chi_+(-k)\right],
\end{equation}
provided that the bulk fermions satisfy their equations of motion. These boundary terms vanish when one imposes Dirichlet boundary conditions on $\chi_+$ and
$\chi_-$ as $\delta\chi_+(r=\epsilon)=\delta\chi_-(r=\epsilon)=0$. 
However, imposing Dirichlet boundary conditions on them simultaneously is clearly impossible
since they are related by equations of motion(e.g. (\ref{manipulation-of-eq-chi+})). In fact, requesting Dirichlet boundary condition to $\chi_+$ is consistent with doing
Neumann boundary condition to $\chi_-$ and vice versa. 
To resolve this problem, one can add a boundary term as
\begin{equation}
S_b=-i\int_{r=\epsilon} d^d k \bar \chi_+(k) \chi_-(-k),
\end{equation}
to the bulk action.
Then, 
\begin{equation}
\delta(\hat S)=\left.\left.-i\int_{r=\epsilon} d^dk \right[ \delta\bar \chi_+(k) \chi_-(-k) - \bar \chi_-(k)\delta \chi_+(-k)\right],
\end{equation}
where
\begin{equation}
\label{modified-bulk-action}
 \hat S=S_{bulk}+S_b.
\end{equation}

Now, it is manifest that one can impose Dirichlet boundary conditions on the fields, both $\chi_+$ and $\bar\chi_+$ consistently, since $\bar\chi_+=\chi^{\dagger}_+\gamma^{\hat 0}$. 
For further discussion, we will demand Dirichlet boundary conditions on $\chi_+$ and $\bar \chi_+$ and do Neumann boundary condition on  $\chi_-$ and $\bar \chi_-$.
$\hat S$ is our starting point and then the total action for HWRG computation is
\begin{equation}
S_{tot}=\hat S+S_B(\chi_+,\bar\chi_+),
\end{equation}
where $S_B$ is boundary effective action. 

$S_B$ depends only on 
the boundary values of $\chi_+$ and $\bar\chi_+$ only. The reason is the following.
Genuine on-shell degrees of freedom are either $\chi_+$ and $\bar \chi_+$ or $\chi_-$ and $\bar \chi_-$ 
because they are related one another by their equations of motion respectively.
Since we impose Dirichlet boundary condition on $\chi_+$ and $\bar \chi_+$ on AdS boundary, 
those become source terms in the generating functional for the boundary CFT. 
The boundary effective action is a generating functional evaluated at $r=\epsilon$ hypersurface and then
it becomes the functional of the boundary sources: $\chi_+$ and $\bar\chi_+$.

By the same token, one can integrate out $\chi_-$ and $\bar\chi_-$ in the bulk action, $S_{bulk}$ to obtain an action only with genuine degrees of freedom: $\chi_+$ and $\bar\chi_+$ as
\begin{equation}
\label{bulk-ac-xhi+}
S_{bulk}=-\int drd^dk \left[ \partial_r \bar\chi_+(k,r)\frac{k_{\mu}\gamma^{\hat \mu}}{|k|^2}\partial_r\chi_+(-k,r) +\bar\chi_+(k,r)\gamma^{\hat \mu}k_\mu\chi_+(-k,r)\right],
\end{equation}
For this, we have used the equations of motion(\ref{eq-chi+}) and (\ref{eq-chibar-}). This bulk action provides the Klein-Gordon type equations of motion for $\chi_+$ and $\bar\chi_+$ and as we will see, 
this form of the bulk action has the same formal form with Fokker-Planck action evaluated in section \ref{Stochastic quantization of massless fermions in AdS}.

\paragraph{Hamilton-Jacobi equation}
The condition that the total action $S_{tot}$ does not depend on the radial cut-off $\epsilon$, by taking derivative of it with respect to that cut-off, provides a Hamilton-Jacobi type 
equation as
\begin{equation}
\label{Hamilton-Jacobi-equation-fermion}
\partial_\epsilon S_B= \int d^dk k_{\mu}
\left[\left( \frac{\overleftarrow\delta S_B}{\delta \chi_+(-k)}\right)\gamma^{\hat\mu} \left(\frac{\overrightarrow\delta S_B}{\delta \bar\chi_+(k)}\right)
- \bar\chi_+(k) \gamma^{\hat \mu} \chi_+(-k) \right],
\end{equation}
where $\frac{\overrightarrow{\delta}}{\delta \bar\chi}$ denotes that the variational operation acts from left and $\frac{\overleftarrow{\delta}}{\delta \bar\chi}$ does that the variational operation acts from right. To derive the Hamilton-Jacobi equation, we have used definitions of canonical momenta of $\chi_+$ and $\bar\chi_+$ as
\begin{eqnarray}
\Pi_+(k)&\equiv& \frac{\overrightarrow \delta \hat S}{\delta \partial_r \bar\chi_+(-k)}=-i\chi_-(k)= \frac{\overrightarrow \delta S_B}{\delta \bar\chi_+(-k)}, \\ \nonumber
\bar \Pi_+(k)
&\equiv& \frac{\overleftarrow \delta \hat S}{\delta \partial_r \chi_+(-k)}=-i\bar\chi_-(k)= \frac{\overleftarrow \delta S_B}{\delta \chi_+(-k)}
\end{eqnarray}
To solve this equation, we assume that the boundary effective action, $S_B$ has a form of
\begin{equation}
S_B=\Lambda(\epsilon)+\left.\left.\int d^d k \right[ \bar J(k,\epsilon)\chi_+(-k) + \bar\chi_+(k) J(-k,\epsilon)+\bar \chi_+(k)D(k,\epsilon)\chi_+(-k)\right],
\end{equation}
where $\Lambda$ is boundary cosmological constant, $\bar J$ and $J$ are 
boundary source terms and $D$ is double trace deformation term. Substitution of the above ansatz to
(\ref{Hamilton-Jacobi-equation-fermion}) leads the following set of equations:
\begin{eqnarray}
\partial_\epsilon \Lambda(\epsilon)&=& \bar J(k,\epsilon)\gamma^{\hat\mu}k_{\mu}J(-k,\epsilon), \\
\partial_\epsilon \bar J(k,\epsilon)&=& \bar J(k,\epsilon)\gamma^{\hat\mu}k_{\mu} D(k,\epsilon), \\
\partial_\epsilon J(-k,\epsilon)&=& D(k,\epsilon)\gamma^{\hat\mu}k_{\mu} J(-k,\epsilon), \\
{\rm and \ \ \ }
\partial_\epsilon D(k,\epsilon)&=& -k_{\mu}\gamma^{\hat\mu}+D(k,\epsilon) k_\mu \gamma^{\hat\mu} D(k,\epsilon).
\end{eqnarray}
We are interested in solving the last equation to get the radial evolution of the double trace operator and its solution
is given by
\begin{eqnarray}
\label{D-solution}
D(k,\epsilon)&=&i\sum_{\alpha}\chi_{\alpha,-}(k,\epsilon) \chi^{-1}_{\alpha,+}(k,\epsilon)=-\sum_{\alpha,\beta}\frac{[k_{\nu}\gamma^{\hat\nu}]_{\alpha\beta}}{|k|^2}\partial_\epsilon \chi_{\beta,+}(k,\epsilon) \chi^{-1}_{\alpha,+}(k,\epsilon), \\ \nonumber
&=&-\sum_\alpha \frac{E_\alpha(k)}{|k|^2}\partial_\epsilon \chi_{\alpha,+}(k,\epsilon) \chi^{-1}_{\alpha,+}(k,\epsilon)
\end{eqnarray}
where $\chi_{\alpha,+}(k,\epsilon)$ is given in Eq(\ref{sol-most-chi+++}),
the inverse of the fermionic field $\chi^{-1}_{\alpha,+}(k,\epsilon)$ is defined by $\chi^{-1}_{\alpha,+}(k,\epsilon)\chi_{\alpha,+}(k,\epsilon) =1$(the spinor index $\alpha$ is not summed), i.e.
\begin{equation}
\chi^{-1}_{\alpha,+}(k,\epsilon)=\frac{1}{N_{\alpha,+}(\cosh(|k|\epsilon)
 +\Delta_{\alpha,+}\sinh(|k|\epsilon))}\langle \alpha|.
\end{equation}

\paragraph{Explicit evaluation of $D(\epsilon,k)$}
The explicit form of the solution of $\chi_+$ is obtained by using the explicit form of the solution(\ref{sol-most-chi+++}). The form of $D(\epsilon,k)$ is given by
\begin{equation}
 -k_\mu \gamma^{\mu}D(\epsilon,k)=
 \sum_{\alpha,\beta}|k|\delta_{\alpha\beta}\frac{\sinh(|k|\epsilon)+\Delta_{\alpha,+}\cosh(|k|\epsilon)}{\cosh(|k|\epsilon)
 +\Delta_{\alpha,+}\sinh(|k|\epsilon)}|\alpha\rangle\langle\beta|,
\end{equation}
or by using Eigen value, $E_\alpha(k)$,
\begin{equation}
\label{Doubkl-tra-couplking}
D_{\alpha\beta}(\epsilon,k)=-
\frac{E_\alpha(k)}{|k|}\delta_{\alpha\beta}\frac{\sinh(|k|\epsilon)
+\Delta_{\alpha,+}\cosh(|k|\epsilon)}{\cosh(|k|\epsilon)+\Delta_{\alpha,+}\sinh(|k|\epsilon)},
\end{equation}

\section{Stochastic quantization of massless fermions in AdS$_{d+1}$}
\label{Stochastic quantization of massless fermions in AdS}

\paragraph{Boundary on-shell action}
We start with the modified bulk action, $\hat S$ defined in (\ref{modified-bulk-action}) 
to obtain the boundary on-shell action. By using the bulk equations of
motion, $\hat S$ becomes a form of
\begin{equation}
\label{pre-on-shell-action}
\hat S=-i\int_{r=\epsilon} d^dk \bar \chi_+(k,r)\chi_-(-k,r),
\end{equation}
which is called on-shell action, $I_{os}$. This on-shell action need to be evaluated from the regular solution of the bulk equations of motion. The solutions 
(\ref{sol-most-chi+}), (\ref{sol-most-chi----}), (\ref{eq--+++}) and (\ref{eq+++--}) are not the regular solutions since they are divergent in the interior of AdS space. To remove the divergences, we 
restrict these solutions by imposing 
\begin{equation}
\chi^{(0)}_\pm+\chi^{(1)}_\pm=0, {\rm\ \ and\ \ } \bar\chi^{(0)}_\pm+\bar\chi^{(1)}_\pm=0.
\end{equation}
Then, the regular solutions are given by
\begin{equation}
\chi_{+}({k,r})=\chi^{(0)}_{+}(k)e^{-|k|r} {\rm \ \ and\ } \chi_{-}(k,r)=-\chi^{(1)}_{-}(k)e^{-|k|r},
\end{equation}
and 
\begin{equation}
\bar\chi_{+}({k,r})=\bar\chi^{(0)}_{+}(k)e^{-|k|r} {\rm \ \ and\ } \bar\chi_{-}(k,r)=-\bar\chi^{(1)}_{-}(k)e^{-|k|r},
\end{equation}
where again we impose Dirichlet B.C to $\chi_+$($\bar\chi_+$) solutions and do Neumann B.C to $\chi_-$($\bar\chi_-$). 
By substituting of the above solutions into the on-shell action(\ref{pre-on-shell-action}), we get
\begin{equation}
 I_{os}=\int d^d k\bar\chi_+(k,\epsilon)\left(\frac{k_\mu \gamma^{\hat \mu}}{|k|} \right)\chi_+(-k,\epsilon),
\end{equation}
where $\chi_+(k,\epsilon)=\chi_+(k)e^{-|k|\epsilon}$ and $\bar\chi_+(k,\epsilon)=\bar\chi_+(k)e^{-|k|\epsilon}$.
We note that $I_{os}$ is written in terms of the boundary value of $\chi_+$ and $\bar\chi_+$ since we impose Dirichlet B.C for them. 

As discussed in \cite{Oh:2012bx,Jatkar:2013uga}, the classical action $S_c$ in stochastic quantization is obtained fro0m the bulk on-shell action with identification as
$S_c=-2I_{os}$. Then, the classical action is
\begin{equation}
\label{the-classical-ActioN}
 S_c=-2\int d^d k\bar\chi_+(k,\epsilon)\left(\frac{k_\mu \gamma^{\hat \mu}}{|k|} \right)\chi_+(-k,\epsilon),
\end{equation}
which becomes the starting point of the stochastic process of the fermionic fields.

\subsection{Langevin approach}
We start with the classical action(\ref{the-classical-ActioN}) to study stochastic quantization of it. Before we start, we note that 
there is a problem for the $naive$ Langevin equations of fermionic fields. The Langevin equations for the fermions may be given by
\begin{equation}
\label{forma-FERMIONIC-Lang-chi}
\frac{\partial \chi(k,t)}{\partial t}=-\frac{1}{2}\frac{\overrightarrow{\delta}S_c}{\delta \bar\chi(-k,t)}+\eta(k,t)
\end{equation}
and
\begin{equation}
\label{forma-FERMIONIC-Lang-chibar}
\frac{\partial \bar\chi(k,t)}{\partial t}=-\frac{1}{2}\frac{\overleftarrow{\delta}S_c}{\delta \chi(-k,t)}+\bar\eta(k,t),
\end{equation}
where $\frac{\overrightarrow{\delta}}{\delta \bar\chi}$ denotes the variational operation acts from left and 
$\frac{\overleftarrow{\delta}}{\delta\chi}$ does the variation acts from right. We also skip the subscript $+$ for the fermionic fields since
we will deal with $\chi_+$ and $\bar\chi_+$ only in this section. The stochastic expectation values of the noise fields $\eta$ and $\bar\eta$ will be given by
\begin{equation}
\langle\eta\rangle=\langle\bar\eta\rangle=0 {\rm \ \ and\ } \langle \eta_\alpha(k,t)\bar\eta_\beta(k^\prime,t^\prime) \rangle =\delta_{\alpha\beta}
\delta(k+k^\prime)\delta(t-t^\prime)
\end{equation}
and N-point correlation functions can be constructed in the similar fashion with the scalar and vector 
fields cases\cite{Oh:2012bx,Jatkar:2013uga,Huffel1} 
by taking into account the anti commuting nature of the Grassmann variables. In fact, the correlations 
of $\eta$ and $\bar\eta$ come from the Gaussian type partition function for the noise fields as
\begin{equation}
Z=\int \mathcal D\eta \mathcal D\bar\eta \exp\left( -\int dtd^dk \delta^{\alpha\beta} \bar\eta_\alpha(k,t^\prime)\eta_\beta(-k,t^\prime)\right).
\end{equation}

Let us evaluate the Langevin equations explicitly by using (\ref{the-classical-ActioN}), which are given by
\begin{equation}
\label{chi-eg}
\frac{\partial \chi(k,t)}{\partial t}=\frac{k_\mu \gamma^{\hat \mu}}{|k|}\chi(k,t)+\eta(k,t)
\end{equation}
and
\begin{equation}
\label{Lange-plus}
\frac{\partial \bar\chi(k,t)}{\partial t}=\bar\chi(k,t)\frac{k_\mu \gamma^{\hat \mu}}{|k|}+\bar\eta(k,t).
\end{equation}
We will discuss the solutions of (\ref{chi-eg}) first. For the solution of Eq(\ref{Lange-plus}), the similar argument with the case of (\ref{chi-eg}) will apply.
The formal solution of (\ref{chi-eg}) has a form of 
\begin{equation}
\label{Naiive-st-solu}
\chi(k,t)=\int^t\exp\left(\frac{k_\mu \gamma^{\hat \mu}(t-t^\prime)}{|k|} \right)\eta(k,t^\prime) dt^\prime,
\end{equation}
where again we point out that it is a formal solution since $\gamma$-matrices cannot be defined as the argument of the exponential function. It will be well posed when the exponential factor is expanded. 

Now, we discuss the convergence of this solution. Since stochastic process will be required to produce its correlation functions correctly 
as it approaches thermal equilibrium, the matrix $k_\mu \gamma^\mu$
in the exponent in Eq(\ref{Naiive-st-solu}) should have negative definite Eigen values for the exponential factor to sufficiently decay as $t\rightarrow \infty$
and the solution converges. However, in general it is not ensured that all 
the Eigen values of the matrix $k_\mu\gamma^{\mu}$ are negative definite. 

For example, let us evaluate the Eigen values of the matrix  $k_\mu\gamma^{\mu}$ in 4-d Euclidean flat spacetime. We use the following representation of $\gamma$-matrices:
\begin{equation}
\gamma^0=\left(\begin{array}{cc}0 & \mathbb I  \\ \mathbb I & 0\end{array}\right) {\rm \ \ and \ }
\gamma^i=i\left(\begin{array}{cc}0 & \mathbb \sigma^i  \\ -\mathbb \sigma^i & 0\end{array}\right).
\end{equation}
It turns out that diagonalization of the matrix  $k_\mu\gamma^{\mu}$ has a form of
\begin{equation}
\label{eigen-values-of-k-gamma}
k_\mu\gamma^{\mu}=\left(\begin{array}{cccc}-\sqrt{k^2} & 0 &0&0  \\ 0 & -\sqrt{k^2} &0&0  \\ 0 & 0 &\sqrt{k^2}&0  \\ 0 & 0 &0&\sqrt{k^2} \end{array}\right),
\end{equation}
where $k^2=k_0^2+k_1^2+k_2^2+k_3^2$. 
The half of the Eigen values are positive definite and they will give divergent solution as $t\rightarrow\infty$.

The cure of this problem is suggested in \cite{Huffel1}. The Langevin equation can be generalized with non trivial kernel as
\begin{equation}
\label{langevin-with-kernel}
\frac{\partial \chi(k,t)}{\partial t}=-\frac{1}{2}V(k)\frac{{\overrightarrow\delta}S_c}{\delta \bar\chi(-k,t)}+\eta(k,t),
\end{equation}
where $V(k)$ is the momentum dependent kernel.
Definition of stochastic correlation functions are also modified to be consistent with the Langevin equation as
\begin{equation}
\label{mofif-2-pt-corr}
\langle\eta\rangle=\langle\bar\eta\rangle=0 {\rm \ \ and\ } \langle \eta_\alpha(k,t)\bar\eta_\beta(k^\prime,t^\prime) \rangle =V(k)_{\alpha\beta}\delta(k+k^\prime)\delta(t-t^\prime),
\end{equation}
which are derived from a Gaussian form of the partition function with the non-trivial kernel as
\begin{equation}
Z=\int \mathcal D\eta \mathcal D\bar\eta \exp\left( -\int dtd^dk  \bar\eta_\alpha(k,t^\prime)[V^{-1}]^{\alpha\beta}(k)\eta_\beta(-k,t^\prime)\right).
\end{equation}
$V^{-1}(k)$ is inverse of the kernel $V(k)$, defined by $V(k)V^{-1}(k)=V^{-1}(k)V(k)=\mathbb I$. 

Thanks to this kernel, the formal solution of (\ref{langevin-with-kernel}) is given by
\begin{equation}
\label{langevin-sol}
\chi(k,t)=\int^t\exp\left(\frac{V(k)k_\mu \gamma^{\hat \mu}(t-t^\prime)}{|k|} \right)\eta(k,t^\prime) dt^\prime.
\end{equation}
Now, one can choose $V(k)$ to be
\begin{equation}
V(k)=-k_{\nu}\gamma^{\hat \nu}f(|k|),
\end{equation}
where $f(k)$ is a positive definite function of momentum $|k|$ and this ensures the convergence of the solution. 

Fixing the form of $f(|k|)$ is another issue and we will make a choice of it with a certain scaling argument which is
consistent with AdS isometry. We have chosen 
\begin{equation}
\label{the-correct-kernel}
f(|k|)=1, {\rm\ i.e.\ \ } V(|k|)=-k_\mu \gamma^{\hat \mu},
\end{equation}
and once we have done this, 
it turns out that the generalized Langevin equation(\ref{langevin-with-kernel}) 
and stochastic 2-point correlation function(\ref{mofif-2-pt-corr}) are invariant under a global scaling symmetry as
\begin{equation}
t \rightarrow \lambda t, k \rightarrow \lambda ^{-1}k, \chi \rightarrow \lambda^{d/2}\chi {\rm \ and\ \ } \eta \rightarrow \lambda^{d/2-1}\eta,
\end{equation}
where $\lambda$ is a constant.
We claim that this is the right choice of the kernel. The reason is the following. In the end, we want identify the stochastic time `$t$'  with the radial variable of AdS space, `$r$' and the momenta along boundary direction $k_\mu$ to those along $d$-dimensional momenta appearing in Langevin equation(or Fokker-Planck action) according to the prescription in \cite{Oh:2012bx,Jatkar:2013uga}. The natural expectation is that the above scaling symmetry is that of bulk fermionic action and when we turn off the fermionic fields($\chi=\eta=0$), then this becomes scaling isometry of AdS space, given by
\begin{equation}
r \rightarrow \lambda r{\rm \ and \ \ } x^\mu \rightarrow \lambda x^\mu,
\end{equation}
where again we note that $k_\mu$ is momenta along $x^\mu$ direction and they scale inversely. 

Once we have chosen the kernel $V(|k|)=-k_\mu\gamma^{\hat \mu}$, the solution(\ref{langevin-sol}) becomes
\begin{equation}
\chi(k,t)=\int^t\exp\left(-|k|(t-t^\prime)\right)\eta(k,t^\prime) dt^\prime.
\end{equation}
and it shows definite convergence as $t\rightarrow \infty$. $\bar\chi$ satisfies the similar equations as
\begin{equation}
\frac{\partial \bar\chi(k,t)}{\partial t}=-\bar\chi(k,t)|k|+\bar\eta(k,t)
\end{equation}
and its solution is given by
\begin{equation}
\bar\chi(k,t)=\int^t\exp\left(-|k|(t-t^\prime)\right)\bar\eta(k,t^\prime) dt^\prime.
\end{equation}

Now, we compute stochastic(equal-time) 2-point correlation function as
\begin{eqnarray}
\langle \chi_{\alpha}(k,t)\bar\chi_\beta(k^\prime,t)\rangle&=&\int^t_{t_0}\int^{t}_{t_0}\exp\left[-|k|(t-\tilde t)-|k^\prime| (t-\tilde t^\prime)\right]\langle \eta_{k,\alpha}(\tilde t)\bar\eta_{k^\prime,\beta}(\tilde t^\prime) \rangle d\tilde t d\tilde t^\prime \\ \nonumber
&=&k_\mu \gamma^{\hat \mu}_{\alpha\beta}\delta(k+k^\prime)\int^t_{t_0}d\tilde t 
\exp[-2|k|(t-\tilde t)]\\ \nonumber
&=&\frac{k_\mu \gamma^{\hat \mu}_{\alpha\beta}}{2|k|}\delta(k+k^\prime)\left( 1-e^{-2|k|(t-t_0)} \right)\\ \nonumber
&=&\frac{E_\alpha(k) \delta_{\alpha\beta}}{2|k|}\delta(k+k^\prime)\left( 1-\frac{\tilde\Delta_\alpha-1}{\tilde\Delta_\alpha+1}e^{-2|k|t} \right),
\end{eqnarray}
where for the last equality, 
we have used Eigen values of the matrix $k_\mu\gamma^\mu$ and chosen the initial time as $t_0=-\frac{1}{|k|}coth^{-1}\tilde\Delta_\alpha$ to match this result with the HWRG computation. Finally, it turns out that the stochastic 2-point correlation function correctly reproduces the double trace coupling of the boundary effective action from the bulk fermionic fields via the following relation
\begin{equation}
\langle \chi_\alpha(k,t) \bar \chi_\beta(k,t) \rangle_H^{-1} = \langle \chi_\alpha(k,t) \bar \chi_\beta(k,t) \rangle_S^{-1}-\frac{1}{2} \frac{\overrightarrow\delta}{\delta \bar \chi_\alpha(-k,t)} 
S_c \frac{\overleftarrow\delta}{\delta \chi_\beta(-k,t)},
\end{equation}
which is the fermionic extension of the relation suggested in \cite{Oh:2012bx,Jatkar:2013uga}, once the stochastic time 
$t$ is identified to the AdS radial coordinate $r$ and $\tilde\Delta_\alpha=\Delta_{\alpha,+}$.

\subsection{Fokker-Planck approach}
To evaluate the form of the Fokker-Planck action, we just start from the stochastic partition function for fermions, given by
\begin{equation}
Z=\int[D\eta D\bar\eta]\exp\left( -\int dt^\prime d^dk \bar \eta_\alpha(k,t^\prime)[V^{-1}]^{\alpha\beta}\eta_\beta(-k,t^\prime)\right).
\end{equation}
To obtain the partition function in terms of the fermionic fields $\chi$ and $\bar\chi$, we replace $\eta$ and $\bar\eta$ with them by using Langevin equations given in Eq(\ref{langevin-with-kernel}). For the manipulation of measure part, we use  Jacobian factor as
\begin{equation}
det\left(\frac{\delta\eta}{\delta\chi}\right)det\left(\frac{\delta\bar\eta}{\delta\bar\chi}\right)
=\exp\left(\int dt^\prime d^d k\frac{\overrightarrow\delta}{\delta \bar\chi(k)}S_c\frac{\overleftarrow\delta}{\delta\bar\chi(-k)}\right).
\end{equation}
Then, the partition function becomes
\begin{equation}
Z=\int[D\chi D\bar\chi]\exp\left[ - S_{FP}
  - \frac{1}{2}S_c(\chi,\bar\chi)\right].
\end{equation}
$S_{FP}$ is called Fokker-Planck action, which is defined by
\begin{equation}
S_{FP}=\int dt^\prime d^dk
\mathcal L_{FP}(\chi,\bar\chi; t^\prime),
\end{equation}
where $\mathcal L_{FP}$ is Fokker-Planck Lagrangian density, given by
\begin{eqnarray}
\mathcal L_{FP}= 
\frac{\partial \bar \chi_\alpha(k,t)}{\partial t} 
[V^{-1}]^{\alpha\beta}(k)\frac{\partial  \chi_\beta(-k,t)}{\partial t}+ \frac{1}{4}\frac{\overleftarrow\delta S_c}{\delta \chi_\alpha}V_{\alpha\beta}(k)\frac{\overrightarrow\delta S_c}{\delta \bar\chi_\beta}-\frac{1}{4}\delta_{\alpha\beta}\frac{\overrightarrow\delta}{\delta \bar\chi_\alpha(k)}S_c\frac{\overleftarrow\delta}{\delta\chi_\beta(-k)}
\end{eqnarray}
The last term in $\mathcal L_{FP}$ has no field dependences at all since we will deal with quadratic Lagrangian only. Therefore, this term will be dropped out.

Once one plugs right forms of the kernel $V(|k|)$ chosen in Eq(\ref{the-correct-kernel}) and $S_c$ given in Eq(\ref{the-classical-ActioN}), then $S_{FP}$ becomes
\begin{equation}
\label{on-shell-fpa}
S_{FP}=-\int dt \left[
\frac{\partial \bar \chi(k,t)}{\partial t} \left(\frac{k_\mu \gamma^{\hat \mu}}{k^2}\right)\frac{\partial  \chi(-k,t)}{\partial t}+\bar\chi(k,t)k_\mu \gamma^{\hat \mu}\chi(-k,t)\right],
\end{equation}
where the spinor indices are suppressed.
We note that this Lagrangian density is the same form with the bulk Lagrangian density(\ref{bulk-ac-xhi+}).

Without considering quantum effects, we will just obtain on-shell Fokker-Planck action.
The equation of motion obtained from Eq(\ref{on-shell-fpa}) is a Klein-Gordon form of equation, which is given by
\begin{equation}
\label{CHI-klein-GORDON}
\partial^2_t \chi_\alpha(k,t)-k^2 \chi_\alpha(k,t)=0.
\end{equation}
In fact, the total derivative term is only left over in the action when we plug the equation of motion into Eq(\ref{on-shell-fpa}). Then, $S_{FP}$ becomes
\begin{equation}
\label{Sfp-boundary}
S_{FP}=-\int d^dk \left(\left. \bar \chi(k,t^\prime)\frac{k_\mu \gamma^{\hat \mu}}{k^2}\frac{\partial \chi(-k,t^\prime)}{\partial t^\prime} \right)\right|^{t^\prime=t}_{t^\prime=t_0}.
\end{equation}

Let us evaluate Eq(\ref{Sfp-boundary}) more in detail by substituting 
the precise form of the solution of the equation of motion(\ref{CHI-klein-GORDON}) into it. 
The most general solution of the equation of motion will be comprised of linear combination of $\sinh(|k|t)$ and $\cosh(|k|t)$.
For the boundary condition at $t^\prime=t$, we want  $\chi(k,t^\prime=t)=\chi(k,t)$. The form of the solution that satisfies this boundary condition is 
\begin{equation}
\chi_\alpha(k,t^\prime)=\zeta^{-1}_\alpha(k,t)\zeta_\alpha(k,t^\prime)\chi_\alpha(k,t)
\end{equation}
where the spinor index $\alpha$ is not summed and
\begin{equation}
\zeta_\alpha(k,t)=\tilde N_\alpha[\cosh(|k|t)+\tilde\Delta_\alpha(k)\sinh(|k|t)]|\alpha\rangle,
\end{equation}
where $\tilde N_\alpha$ is normalization constant and $\tilde\Delta(k)_\alpha$ is an arbitrary momentum dependent function. We can obtain the solution for $\bar\chi$ in the similar fashion as
\begin{equation}
\bar\chi_\alpha(k,t^\prime)=\bar\chi_\alpha(k,t)\bar\zeta_\alpha(k,t)\bar\zeta^{-1}_\alpha(k,t^\prime),
\end{equation}
where the index $\alpha$ is not summed,  
$\bar\zeta_\alpha$ is the solution of the equation of motion derived from the Fokker-Planck Lagrangian, which is given by
\begin{equation}
\bar\zeta_\alpha(k,t)=\bar\zeta_{\alpha,0}(k)\cosh(|k|t)+\bar\zeta_{\alpha,1}(k)\sinh(|k|t),
\end{equation}
where $\bar\zeta_{\alpha,0}(k)$ and $\bar\zeta_{\alpha,1}(k)$ are arbitrary momentum dependent spinors.
$\bar \zeta^{-1}(k,t)$ is a spinor satisfying the condition as $\bar \zeta_\alpha(k,t)\bar \zeta^{-1}_\alpha(k,t)=1$, where the spinor index $\alpha$ is not summed.

Finally, we get
\begin{eqnarray}
S_{FP}&=&\sum_{\alpha,\beta}\int d^dk\left[ \bar\chi_\alpha(k,t)\left(\frac{-[k_\mu\gamma^{\hat\mu}]_{\alpha\beta}}{|k|^2}\right)\zeta_\beta^{-1}(-k,t)\frac{\partial \zeta_\beta(-k,t)}{\partial t}\chi_\beta(-k,t) \right], \\ \nonumber
&=&\sum_\alpha\int d^dk\left[ \bar\chi_\alpha(k,t)\left(\frac{-E(k)\delta_{\alpha\beta}}{|k|^2}\zeta_\beta^{-1}(-k,t)\frac{\partial \zeta_\beta(-k,t)}{\partial t}\right)\chi_\beta(-k,t) \right], \\ \nonumber
&=&\sum_\alpha\int d^dk\left[ \bar\chi_\alpha(k,t)\left(\frac{-E(k)\delta_{\alpha\beta}}{|k|}\frac{\sinh(|k|t)+\tilde\Delta_\alpha\cosh(|k|t)}{\cosh(|k|t)+\tilde\Delta_\alpha\sinh(|k|t)}\right)\chi_\beta(-k,t) \right], \\ \nonumber
&\equiv&\int d^dk  \bar\chi(k,t)\mathcal D(k,t)\chi(-k,t),
\end{eqnarray}
where $\mathcal D(k,t)$ is precisely the same with the double trace coupling $D(k,r)$, given in Eq(\ref{Doubkl-tra-couplking}) once the AdS radial coordinate $r$ is identified to the stochastic time $t$ and $\tilde\Delta_\alpha(k)$ is done to
$\Delta_{\alpha,+}(k)$.

We point out that we have used the same initial boundary condition that we have requested to the solutions of Langevin equation.
For the boundary condition for the initial time $t_0$, 
\begin{equation}
t_0=-\frac{1}{|k|}coth^{-1} \tilde\Delta_\alpha,
\end{equation}
which removes any contributions at $t=t_0$ in $S_{FP}$.

\section*{Acknowledgement}
J.H.O would like to thank his $\mathcal W.J$. He also thank Dileep P. Jatkar for many useful discussions.
This work is supported by the research fund of Hanyang University(HY-2013) only.



\end{document}